\documentclass[11pt,floatfix,altaffilletter,superscriptaddress,tightenlines,showpacs,showkeys,preprintnumbers,nofootinbib]{revtex4-1}
\pdfoutput=1
\usepackage[colorlinks=true,citecolor=blue,linkcolor=blue,breaklinks=true]{hyperref}
\usepackage{amsmath,amssymb}
\usepackage{epsfig} 
\usepackage{graphicx}        
\usepackage{url}
\usepackage{color}
\usepackage{multirow}
\usepackage{placeins}
\usepackage[dvipsnames]{xcolor}
\usepackage{braket}
\usepackage{float}
\usepackage{slashed}
\usepackage{fdsymbol}
\hypersetup{
  colorlinks=true,
  linkcolor=red,
  filecolor=magenta,   
  urlcolor=blue,
  citecolor=blue,
}

\allowdisplaybreaks

\bibpunct{[}{]}{,}{n}{}{,}

\def\beq{\begin{equation}}
\def\eeq{\end{equation}}
\def\bea{\begin{eqnarray}}
\def\eea{\end{eqnarray}}
\makeatletter
\@addtoreset{equation}{section}

\begin{document}
%\title{Probing Type-I seesaw with curvature-dependent lepton chemical potential}
%\title{Flavour effects in gravitational leptogenesis}
%\title{Importance of flavour effects in right  handed neutrino induced gravitational leptogenesis}
\title{ Probing Leptogenesis and Pre-BBN Universe with Gravitational Waves Spectral Shapes}
%\author{Madhurima Pandey}
%\email{madhurima.pandey@saha.ac.in}
%\affiliation{Astroparticle Physics and Cosmology Division, Saha Institute of Nuclear Physics, HBNI 1/AF Bidhannagar, Kolkata 700064, India}
%\author{Avik Paul}
%\email{avik.paul@saha.ac.in}
%\affiliation{Astroparticle Physics and Cosmology Division, Saha Institute of Nuclear Physics, HBNI 1/AF Bidhannagar, Kolkata 700064, India}
\author{Rome Samanta}
\email{romesamanta@gmail.com}
\affiliation{CEICO, Institute of Physics of the Czech Academy of Sciences, Na Slovance 1999/2, 182 21 Prague 8, Czech Republic}
\author{Satyabrata Datta}
\email{satyabrata.datta@saha.ac.in}
\affiliation{Saha Institute of Nuclear Physics, HBNI, 1/AF Bidhannagar,
Kolkata 700064, India}

%\preprint{NUHEP-TH/19-08}

\begin{abstract} 
On the frequency-amplitude plane, Gravitational Waves (GWs) from cosmic strings show a flat plateau at higher frequencies due to the string loop dynamics in standard radiation dominated post-inflationary epoch. The spectrum may show an abrupt upward or a downward trend beyond a turning point frequency $f_*$, if the primordial dark age prior to the Big Bang Nucleosynthesis (BBN), exhibits non-standard cosmic histories.  We argue that such a spectral break followed by a rising GW amplitude which is a consequence of a post-inflationary equation of state ($\omega>1/3$) stiffer than the radiation ($\omega=1/3$), could also be a strong hint of a leptogenesis in the seesaw model of neutrino masses. Dynamical generation of the right handed (RH) neutrino masses by a gauged $U(1)$ symmetry breaking leads to the formation of a network of cosmic strings which emits stochastic GWs. A gravitational interaction of the lepton current by an operator of the form $\partial_\mu R j^\mu$--which can be generated in the seesaw model at the two-loop level through RH neutrino mediation, naturally seeks a stiffer equation of state to efficiently produce  baryon asymmetry  proportional to $1-3\omega$. We discuss how GWs with  reasonably strong amplitudes complemented by a neutrino-less double beta decay signal  could  probe the onset of the most recent radiation domination and lightest RH neutrino mass at the  intermediate scales.
\end{abstract}

\maketitle

\section{Introduction}
Leptogenesis\cite{lep1,lep2,lep3,lep4,lep5,lep6,lep7} is a simple mechanism to explain the observed baryon asymmetry of the universe\cite{Planck}. The right handed (RH) heavy neutrinos which are introduced in the Standard Model (SM) to generate light neutrino masses (Type-I seesaw), decay CP asymmetrically to create lepton asymmetry which is then converted to baryon asymmetry via Sphaleron transition\cite{Kuzmin:1985mm}. When it comes to the testability of leptogenesis, there are subtleties. If the heavy neutrino masses are not protected by any symmetry\cite{Altarelli:2010gt},  it is quite natural to assume that they are hierarchical in nature like any other family of SM fermions. In that case, the lightest RH mass scale is bounded from below $M\gtrsim 10^9$ GeV\cite{Davidson:2002qv}which is  beyond the reach of the present collider experiments. Nonetheless, still the colliders and other low energy neutrino experiments  can probe leptogenesis mechanisms that do not constitute hierarchical RH neutrinos--starting from $\mathcal{O}$(TeV) to $\mathcal{O}$(MeV) scale heavy neutrinos\cite{Ars,Ham,Pila,dev1}. A shift of attention from the collider experiments to the Gravitational Waves (GWs) physics is not less interesting in terms of testing leptogenesis. Particularly,  this new cosmic frontier, in which after the discovery of GWs from black hole mergers by LIGO and Virgo collaboration\cite{gw1,gw2}, plenty of efforts are being made to detect primordial GWs from the Early Universe (EU) within a wide range of frequencies--starting from the Pulsar Timing Arrays (PTAs, $\sim$ nHz) to the LIGO($\sim$ 25Hz).

 A network of cosmic strings\cite{cs1,cs2,cs3} which is a generic consequence of  breaking symmetries such as  $U(1)$, is one of the prominent sources of strong stochastic primordial gravitational waves which can be tested in a complementary way in most of the planned GW detectors. Numerical simulations based on the Nambu–Goto action\cite{ng1,ng2} indicate that cosmic string loops loose energy dominantly via GW radiation, if the underlying broken symmetry corresponds to a local gauge symmetry. In the context of seesaw, this sounds music to the ears, since such a gauge symmetry is $U(1)_{B-L}$\cite{Davidson:1978pm,moha1,moha2}, breaking of  which could be responsible for the dynamical generation of the heavy RH masses and hence the lepton number violation as well as creation of  a network of cosmic strings. Having this set-up, there could be two categories to look for the GWs as a probe of leptogenesis. {\it Category A:} A scale separation between the RH masses and the typical Grand Unified Theory (GUT) scale ($\sim 10^{16}$ GeV), imposed by seesaw perturbativity condition  and the neutrino oscillation data\cite{nuosc} implies that residual symmetries like $U(1)_{B-L}$ protects the RH neutrinos to get mass at the GUT scale. Therefore, breaking of that symmetry at a later  stage and consequent emission of GWs from cosmic strings are  natural probes of the scale of leptogenesis. In this case, it is the amplitude (GW energy density normalised by the critical energy density) of the GWs that matters as a probe  and this approach has been taken in Refs.\cite{lepcs1,lepcs2}. {\it Category B:} To make the testability more robust, along with the amplitudes, one can associate leptogenesis also to the spectral shapes of the GWs\cite{lepcs3,lepcs4}. Cosmic string loops that originate and decay in the radiation domination, exhibit a flat plateau on the amplitude-frequency plane at the higher frequencies. This spectral shape may show an  upward or a downward trend if something other than radiation dominates the energy density of the EU before the onset ($T_*$) of most recent radiation domination prior to the BBN ($T\sim 5$ MeV)\cite{bbn,bbn1,bbn2}. Such a non-standard cosmic history that is  responsible for this spectral break which along with the GW amplitude, one aims to claim also as a probe, should therefore be a natural/well-motivated call from the perspective of leptogenesis. Two well-known scenarios in this context can be opted for. {\it Category B1:} A matter domination ($\omega=0<1/3$)\cite{matdom1,matdom2}. {\it Category B2:} Scenarios such as kination ($\omega=1>1/3$)\citep{kin1,kin2}. For the former (latter), one finds a spectral break followed by  a downward (upward) going  GW amplitude\cite{spb0,spb1,spb2,spb3}. Two  leptogenesis mechanisms in the {\it Category B1}--a low-scale leptogenesis and a leptogenesis from ultralight  primordial black holes ($M_{PBH} \lesssim 13g$)  have been studied in Ref.\cite{lepcs3} and Ref.\cite{lepcs4} respectively. In this article, we discuss a scenario that falls in the {\it Category B2}, i.e., interpreting a flat then a spectral break followed by a rising GW amplitude as a signature of leptogenesis. 
 
 Note that, two crucial ingredients for this typical signal are of course cosmic string network itself and then a non-standard equation of state ($\omega=1$ in our discussion).   In the context of leptogenesis from decays\cite{lep1}, though the former is a natural consequence in the sense of  {\it Category A}\cite{lepcs1}, a stiffer equation of state is not an indispensable criterion. However, in seesaw models, even when the  Lagrangian is minimally coupled to gravity, through massive RH neutrino mediation one can generate an operator of the form  $\partial_\mu R j^\mu/M^2$ at two-loop level\cite{grav1,grav2,grav3} (see also Ref.\cite{lepcs2,grav4} for a flavour generalisation and Ref.\cite{grav5} for a recent  review), where $R$ is the Ricci scalar and  $j^\mu$ is the lepton current. This operator is a well-studied operator\cite{grav6,grav7,Mohanty:2005ud,grav8} with the corresponding mechanism dubbed as ``gravitational lepto/baryogenesis" and produces final baryon asymmetry proportional to $\dot{R}\propto (1-3\omega)$. Interestingly, note now that two primary ingredients of the GW signal are also natural requirements to obtain non-zero lepton asymmetry, i.e., the symmetry breaking which gives rise to massive RH neutrinos (mediate in the loops\cite{grav3}) as well as  cosmic strings and then an equation of state $\omega\neq 1/3$\cite{wsm}. We shall discuss later on, that indeed a stiffer equation of state is needed to efficiently produce lepton asymmetry. Plateau amplitudes corresponding to $G\mu\lesssim 10^{-12}$ with $G$ being the Newton constant and $\mu$ being the string tension, with a post LISA spectral break supplemented by a potential test in neutrino-less double beta decay experiments, make the scenario generally robust. The above introduction summarises the basic idea and the main results of this paper. The next sections are dedicated to a more detailed description and technicalities.

\section{gravitational waves from cosmic strings}
Cosmic strings may originate as the fundamental or composite objects in string theory\cite{cssu1,cssu2} as well as topological defects from spontaneous symmetry breaking (SSB) when the vacuum manifold $\mathcal{M}$ has a non-trivial first homotopy group $\pi_1(\mathcal{M})$. A theory with spontaneous breaking of a $U(1)$ symmetry exhibits string solution\cite{cs2,cs3}, since $\pi_1(\mathcal{M})=\mathbb{Z}$. An example of a field theory containing string like solution is a theory of $U(1)$-charged complex scalar field $\phi$ that in the context of seesaw could be a SM scalar singlet $\phi_{B-L}$ which is responsible for the dynamical generation of RH neutrino masses. After the formation, strings get randomly distributed in space and form a network of horizon-size  long strings\cite{csrev1,csrev2} characterised by a correlation length $L=\sqrt{\mu/\rho_\infty}$, where $\mu $--the string tension or energy per unit length is in general constant (however, e.g., in case of global strings\cite{Chang:2021afa} and recently introduced melting strings\cite{Emond:2021vts} $\mu\sim f(T)$) and typically taken to be the square of the symmetry breaking scale $\Lambda_{CS}$ and $\rho_\infty$ is the long string energy density. When two segments of long strings cross each other they inter-commute and form loops with a probability $P=1$\cite{incom1} (exceptions\cite{incom2}). A string network may interact strongly with thermal plasma and thereby its motion gets damped\cite{fric}. After the damping stops, the strings oscillate and enter a phase of scaling evolution that constitute two competing dynamics namely the stretching of the correlation length due to the cosmic expansion and fragmentation of the long strings into loops which oscillate independently and produce particle radiation or gravitational waves\cite{looprad1,looprad2,looprad3}. Out of these two competing dynamics, there is an attractor solution called the scaling regime\cite{scl1,scl2,scl3} in which the characteristic length scales as $L\sim t$. This  implies, for constant string tension, $\rho_\infty\propto t^{-2}$. Therefore, the network tracks any cosmological background energy density $\rho_{bg}\propto a^{-3(1+\omega)}\propto t^{-2}$ with the same equation of state  and hence cosmic strings do not dominate the energy density of the universe like any other defects. The loops radiate GWs at a constant rate which sets up the time evolution of  a loop of initial size $l_i=\alpha t_i$ as $l(\tilde{t})=\alpha t_i-\Gamma G\mu(\tilde{t}-t_i)$, where $\Gamma\simeq 50$\cite{looprad1,looprad3} and the initial loops size parameter $\alpha\simeq 0.1$--a value preferred by numerical simulations\cite{nusim1,nusim2}.  The total energy loss from a loop is decomposed into a set of normal-mode oscillations with frequencies $f_k=2k/l=a(t_0)/a(\tilde{t})f$, where $k=1,2,3...k_{max}$ ($k_{max}$ is for numerical purpose, otherwise $\infty$) and $f$ is the frequency observed today. Given the loop number density $n\left(\tilde{t},l_k\right)$, the present time gravitational wave density parameter is given by $\Omega_{GW}(t_0,f)\equiv f\rho_c^{-1}d\rho_{GW}/df=\sum_k\Omega_{GW}^{(k)}(t_0,f)$, with the $k$th mode amplitude $\Omega_{GW}^{(k)}(t_0,f)$ as\cite{nusim1} 
\bea
\Omega_{GW}^{(k)}(f)=\frac{2kG\mu^2 \Gamma_k}{f\rho_c}\int_{t_{osc}}^{t_0} \left[\frac{a(\tilde{t})}{a(t_0)}\right]^5 n\left(\tilde{t},l_k\right)d\tilde{t}.\label{gwf1}
\eea
The quantity $\Gamma_k$ depends on the small scale structures of the loop and is given by $\Gamma^{(k)}=\frac{\Gamma k^{-\delta}}{\zeta(\delta)}$, e.g.,  $\delta=4/3$ and $5/3$ for cusps and kinks\cite{cuki}. The integration in Eq.\ref{gwf1}  is subjected to a Heaviside function $\Theta\equiv \Theta(t_i-t_{osc})\Theta(t_i-\frac{l_{cric}}{\alpha})$, with $t_{osc}= {\rm Max}~\left[{\rm network~formation~time}(t_F),{\rm end~of~damping (t_{fric})}\right]$ and $l_{cric}$ is the critical length below which massive particle radiation dominates over GWs\cite{partrad1,partrad2}. Both these $\Theta$ functions set a high-frequency cut-off in the spectrum (a systematic analysis can be found in Ref.\cite{spb2}). 

The most important aspect to obtain the GW spectrum is the computation of the loop number density $n\left(\tilde{t},l_k\right)$ which we calculate from the Velocity-dependent-One-Scale (VOS) model\cite{vos1,vos2,vos3} which assumes the loop production function to be a delta function, i.e. all the loops are created with the same fraction of the horizon size with a fixed value of $\alpha$. Given a general equation of state parameter  $\omega$, the number density $n_\omega\left(\tilde{t},l_k\right)$ is computed as
\bea
n_\omega(\tilde{t},l_{k}(\tilde{t}))=\frac{A_\beta}{\alpha}\frac{(\alpha+\Gamma G \mu)^{3(1-\beta)}}{\left[l_k(\tilde{t})+\Gamma G \mu\tilde{t}\right]^{4-3\beta}\tilde{t}^{3\beta}},\label{genn0}
\eea
where $\beta=2/3(1+\omega)$ and we assume $A_\beta =29.6~(\omega=1),5.4~(w=1/3)$ and $0.39~(\omega=0)$\cite{vos3} is a step-function  while changing the cosmological epochs. The most interesting feature of GWs from cosmic string is that the amplitude increases with the symmetry breaking scale $\Lambda_{CS}$. This can be seen by computing the $\Omega_{GW}^{(1)}$, considering loop production as well as decay in the radiation domination which gives an expression for a flat plateau at higher frequencies (see AUX A for an exact formula)
\bea
\Omega_{GW}^{(1)}(f)=\frac{128\pi G\mu}{9\zeta(\delta)}\frac{A_r}{\epsilon_r}\Omega_r\left[(1+\epsilon_r)^{3/2}-1\right], \label{flp1}
\eea
where $\epsilon_r=\alpha/\Gamma G\mu$ and $\Omega_r\simeq 9\times 10^{-5}$. Such strong GWs as a consequence of a very high scale symmetry breaking thus serves as an outstanding probe of particle physics models\cite{pmcs1,pmcs2,pmcs3,pmcs4,pmcs5,pmcs6,Bian:2021vmi}. Possibly the most important recent development is the finding of a stochastic common spectrum process across 45 pulsars by NANOGrav PTA\cite{NANOGrav:2020bcs},  which if interpreted as GWs, corresponds to a strong amplitude and is better fitted with cosmic strings\cite{lepcs2,csfit1,csfit2} than the  single value power spectral density as predicted by supermassive black hole models. Let's also mention that a very recent analysis by PPTA\cite{ppta} is in agreement with the NANOGrav result. In presence of an additional early kination era, the entire GW spectrum is determined by four dynamics. I) A peak at a lower frequency--caused by the loops which are produced in the radiation era and decay in the standard matter era. II) The flat plateau, $\Omega_{GW}^{\rm plt}$, as mention while describing Eq.\ref{flp1}. III) A spectral break at $f_*=\sqrt{\frac{8}{\alpha\Gamma G\mu}}t_*^{-1/2}t_0^{-2/3}t_{\rm eq}^{1/6}$--so called the turning point frequency\cite{spb1,spb2,lepcs4}, followed by a rising GW amplitude $\Omega_{GW}^{(1)}(f>f_*)\simeq \Omega_{GW}^{\rm plt}\left(f/f_*\right) $, caused by modified redshifting of the GWs during kination era V) a second turning point frequency $f_\Delta$ after which the GWs amplitude falls, e.g,  due to particle productions below $l<l_{cric}=\beta_m\frac{\mu^{-1/2}}{(\Gamma G\mu)^m}$, with $\beta_m\sim $ $\mathcal{O}(1)$ and $m=1,2$ for loops with kinks or cusps\cite{partrad1,partrad2}. If the falling is caused due to thermal friction, then one needs to consider the damping of the smaller loops along with the long-string network for $t<t_{fric}$, discarding any GWs production by the smaller loops, i.e, the entire dynamics is completely frozen until $t_{fric}$\cite{fric}. In fact, in our computation we do not take into account any GWs produced from smaller loops prior to $t_{fric}$ and consider that the falling  is due to  particle production which sets the high-frequency cut-off that is much more stronger (appears at lower frequencies) than the friction cut-off\cite{spb2}.  Note also that if the two turning-point frequencies are close to each other,  potentially the GW detectors could see a small bump after the flat plateau with a peak amplitude $\simeq \Omega_{GW}^{\rm plt}\left(f_\Delta/f_*\right)$. Nevertheless, as we show in the next section that given a successful leptogenesis, the second turning point frequency as well as small bumps are most likely to be outside the frequency range of the  GW detectors.

 Before concluding the section, we note two important points. Firstly, the VOS model overestimates the number density of the loops by an order of magnitude  compared to the numerical simulations\cite{nusim1}. This is due to the fact that VOS model considers all the loops are of same size at the production. However, there could be a distribution of $\alpha$. Numerical simulation finds that only 10$\%$ of the energy of the long-string network goes to the large loops ($\alpha\simeq 0.1$) while the rest $90\%$ goes to the highly boosted smaller loops that do not contribute to the GWs. This fact is taken into account by including a normalisation factor $\mathcal{F}_\alpha\sim 0.1$ in Eq.\ref{genn0}\cite{vos3}. Secondly, the amplitude beyond $f_*$ goes as $f^1$ even after taking into account high-$k$ modes (see AUX A) unlike the case of an early matter domination where the same changes from $f^{-1}\rightarrow f^{-1/3}$ for cusps like structures\cite{lepcs3,lepcs4,spb2}.
\section{Gravitational leptogenesis, results and discussion}
The idea behind gravitational leptogenesis\cite{grav7} is, a C and CP-violating operator  $\mathcal{L}_{CPV}\sim b\partial_\mu R j^\mu\sim b\partial_\mu R\bar{\ell}\gamma^\mu\ell$ with $b$ as a real effective coupling, corresponds to a chemical potential $\mu=b\dot{R}$ for the lepton number in the theory.  Therefore, the normalised (by photon density $n_\gamma\sim T^3$) equilibrium lepton asymmetry (using standard Fermi-Dirac statistics with energies $E_\pm=E\pm\mu$) is given by $N_{B-L}^{eq}\sim \frac{b\dot{R}}{T}$. Interestingly,  $\mathcal{L}_{CPV}$ can be generated in a UV framework using the seesaw Lagrangian even when it is minimally coupled to gravity (see e.g., Ref.\cite{grav2} for an in-depth discussion, sec.II of Ref.\cite{lepcs2} for a brief summary). As a computational insight, one calculates  an effective $\ell\ell h$ vertex corresponding to the operator $\mathcal{L}_{CPV}$ using a conformally flat metric $g_{\mu\nu}=(1+h)\eta_{\mu\nu}$ with $R=-3\partial^2h$, capitalising the fact that the coupling `$b$' is independent of the choice of background. In seesaw model, a similar  $\ell\ell h$ vertex that manifests the $\mathcal{L}_{CPV}$ operator, can be constructed  at two-loop level, where the Higgs and the RH masses mediate the loops. Then simply comparing the coefficients of both the vertices up to linear order in $h$,  the coupling $b$ can be calculated in terms of the Yukawa coupling $f$ (where, $f_{\alpha i}\bar{\ell}_{L\alpha}\tilde{H} N_{Ri}$ is the Yukawa interaction in seesaw, with $\ell_{L\alpha}$, $H$ and $N_R$ being the lepton doublet, Higgs and RH fields respectively) and RH neutrino masses $M_i$. The expression for the equilibrium asymmetry then reads 
\bea
N_{B-L}^{eq}=\frac{\pi^2\dot{R}}{36 (4\pi )^4}\sum_{j>i}\frac{{\rm Im}\left[k_{ij}^2\right]}{\zeta (3)T M_i M_j}{\rm ln }\left(\frac{M_j^2}{M_i^2}\right),\label{gl1}
\eea
where $k_{ij}=(f^\dagger f)_{ij}$. The above expression could be modulated by a factor $(M_j^2/M_i^2)^\gamma$, where $\gamma=0,1$. However, $\gamma=0$ appears to be the most natural solution which can be calculated exactly\cite{grav2,grav3}. In any case, even if one considers $\gamma=1$ or the `hierarchical enhancement', tuning the complex part in $k_{ij}^2$, correct baryon asymmetry can always be reproduced. The most important part is, $N_{B-L}\propto \dot{R}\propto 1-3\omega$ which is still vanishing in radiation domination at high temperatures with SM-QCD thermodynamic potential\cite{wsm}. Therefore, a general cosmological background other than radiation that is quite a natural call now, always stems a non-vanishing equilibrium  asymmetry unless the Yukawa couplings are  real or purely imaginary. In the EU, any dynamically produced lepton asymmetry tracks the $N_{B-L}^{eq}$ if the interaction that causes the asymmetry production is strong enough. When the interaction rate becomes weaker (compared to the Hubble expansion), the asymmetry freezes out with the  potential to reproduce correct baryon asymmetry $N_{B-L}\sim 6\times 10^{-8}$\cite{Planck}. In seesaw model, $\Delta L=2$ interactions\cite{dell2} play this role. The general evolution equation that governs the entire dynamics is given by 
\bea
\frac{{ d  N_{B-L}}}{ dz}=-\left(\frac{\kappa}{z^p}+W_{\rm ID}\right)\left[N_{B-L}-\frac{\beta}{z^q}\right],\label{beg0}
\eea
where $z=M_1/T$, $W_{\Delta L=2}(z)=\frac{\kappa}{z^p}$ with $p=\frac{5-3\omega}{2}$, $N_{B-L}^{eq}=\frac{\beta}{z^q}$ with $q=\frac{7+9 \omega}{2}$ and $W_{\rm ID}$ represents the inverse decay $\ell H\rightarrow N_1$ rate.  The parameters $\kappa\sim f_\kappa(m_i,M_1)z_*^{\frac{1}{2}(1-3\omega)}$ and $N_{B-L}^{eq}\propto\beta\sim f_\beta(m_i,M_1, {\rm Im}[f_{ij}])(1-3\omega)z_*^{\frac{3}{2}(3\omega-1)}$, where $z_*=M_1/T_*$ and $m_i$ is the $i$-th light neutrino mass with $i=1,2,3$. All the exact expressions can be found in AUX B. Before proceeding further, let us mention that we do not include the charged lepton flavour effects in this analysis for simplicity. Nonetheless, a systematic description with flavour issues can be found in Ref.\cite{grav4} along with a more finer description in Ref.\cite{lepcs2}. To proceed further, the process consists of two distinct temperature regimes. At a higher temperature $T_{in}\sim \Lambda_{CS}$, as soon as the symmetry breaks, the RH neutrinos become massive and Eq.\ref{beg0} starts acting without $W_{\rm ID}$ which is negligible at this regime. In this gravitational leptogenesis scenario, typically, $z_{in}(=M_1/T_{in})$ can be constrained with so called weak field condition as $z_{in}\ge\sqrt{M_1/\tilde{M}_{Pl}}$, where $\tilde{M}_{Pl}$ is the reduced Planck constant. Once the asymmetry freezes out,  at the lower temperatures,  it faces a washout by the inverse decays which are strongly active at $T\sim M_1$. The final asymmetry is therefore of the form $N_{B-L}^f=N_{B-L}^{G0}e^{-\int_0^\infty W_{\rm ID}  (z) dz}$, where $N_{B-L}^{G0}$ is the frozen out asymmetry after the system is done with $\Delta L=2$ interaction, and the exponential term represents a  late-time  washout by the inverse decays. A general solution of Eq.\ref{beg0} is complicated and depends on the properties of {\it incomplete Gamma functions}. However,  for $\omega=1$, that corresponds to $p=1$ and $q=8$, a simpler solution can be obtained. 
\begin{figure}
\includegraphics[scale=.42]{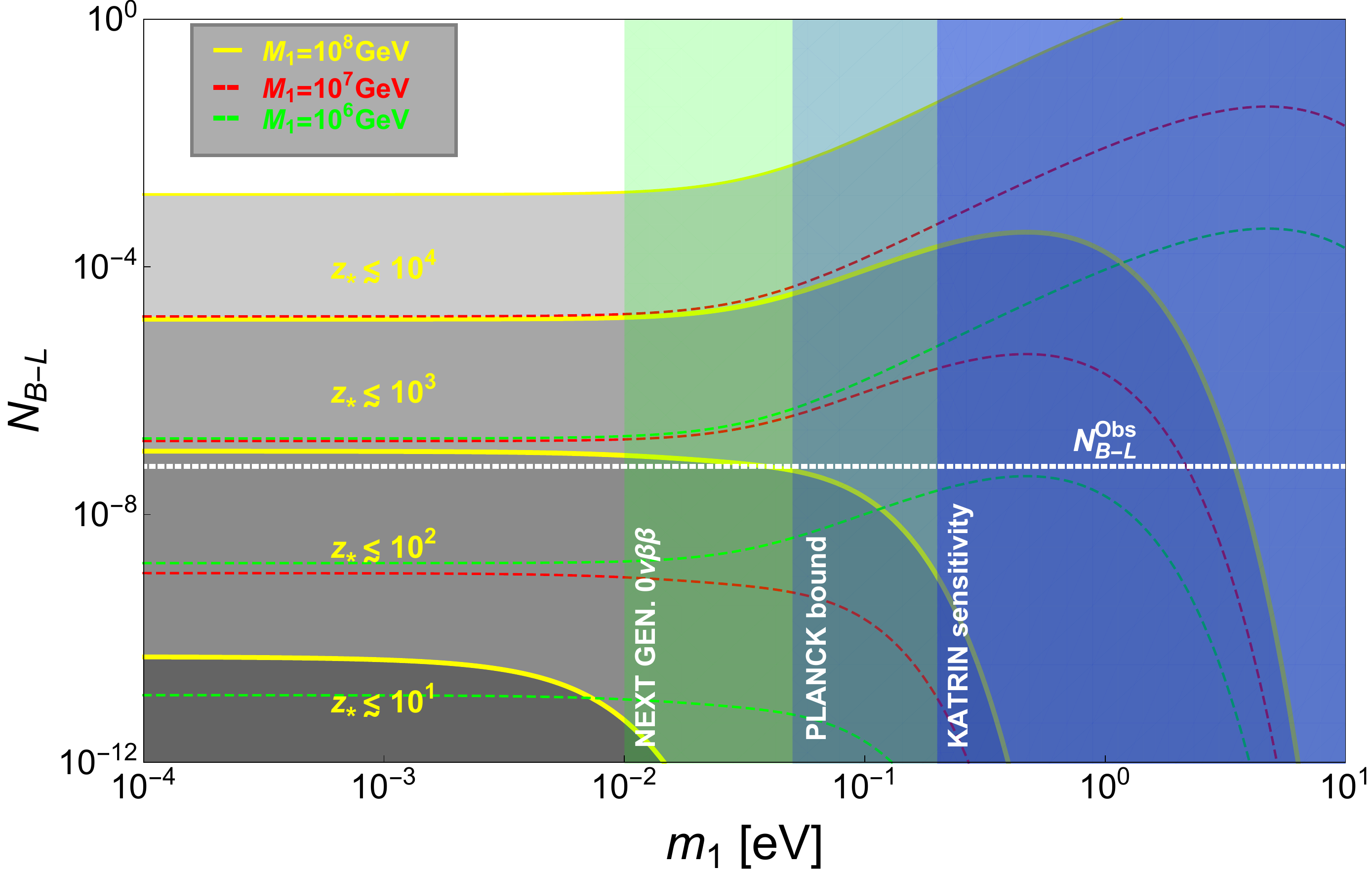}
\caption{Eos: $\omega=1$. The yellow, red and green lines correspond to the lightest RH mass $M_1=10^{8,7,6}$ GeV. For $M_1=10^{7,6}$ we do not show the lines corresponding to $z_*=10^1$. We take $M_3=M_1/z_{\rm in}$ GeV, $M_2=10^{-1}M_3$ GeV, $x_{ij}=\pi/4$, $y_{ij}=10^{-1}$ and two mass-squared differences are at their best-fit values. }\label{fig1}
\end{figure}
We find the  expression for the final asymmetry $N_{B-L}^f(z\rightarrow \infty)$ to be  
\bea
N_{B-L}^f\simeq\frac{\kappa\beta}{8z_{\rm in}^8}{\rm Exp}\left[-\frac{4K_1}{z_*}\right],\label{master}
\eea
where the dimensionless  washout/decay parameter  $K_1$ is a function of Yukawa couplings. Eq.\ref{master} that matches with the numerical solutions of the Eq.\ref{beg0} with quite a high accuracy, is the master equation which we use to present all the results. 

Prior to the explanation of Fig.\ref{fig1}, let's introduce a parametrisation of the Yukawa matrix as $m_D=U\sqrt{m}\Omega\sqrt{M}$, where $m_D=fv$ with $v=174$ GeV, $U$ is the leptonic mixing matrix and $\Omega$ is a $3\times 3$ complex orthogonal matrix with a standard parametrisation in terms of three complex rotation matrices\cite{lepcs2} with complex angles $\theta_{ij}=x_{ij}+i y_{ij}$. In general, $\Omega$ is a completely `free' matrix unless one invokes additional symmetries to fix the flavour structure of the theory. A plethora of works is dedicated in this direction\cite{Altarelli:2010gt}. With this orthogonal parametrisation it is easy to show that the equilibrium asymmetry is independent of $U$. Therefore, as far as the seesaw parameters are concerned, the light, heavy neutrino masses and the orthogonal matrix take part in the process. The decay parameter can also be expressed in terms of these parameters as $K_1=m_*^{-1}\sum_k m_k|\Omega_{k1}|^2$ with $m_*\simeq 10^{-3}$ being the equilibrium neutrino mass\cite{lep4}. In Fig.\ref{fig1}, we show the variation of the produced asymmetry with the lightest neutrino mass for three benchmark values; $M_1=10^{6,7,8}$ GeV with a fixed orthogonal matrix and different values of $z_*$. The basic nature of the curves is quite interesting. Let's focus on the  $z_*=10^3$  curve (yellow) for $M_1=10^8$ GeV. It shows a plateau until $m_1\simeq 10^{-2}$ eV, then an increase followed by a downfall at large $m_1$ values. First of all, for $w=1$, the parameter $\kappa\sim z_*^{-1}$ and therefore for large values of $z_*$, the strength of the $\Delta L=2$ process becomes so weak that the asymmetry instantly freezes out without tracking the equilibrium number density. The coefficient $f_\kappa$ does not change much until $m_1\sim 10^{-2}$ eV and then increases for $m_1\gtrsim 10^{-2}$ eV\cite{lepcs2}. This increase in $f_\kappa$ pushes the asymmetry more towards the equilibrium and hence the overall magnitude of $N_{B-L}$ increases for $m_1\gtrsim 10^{-2}$ eV. A downfall at large $m_1$ is caused by the exponential term in Eq.\ref{master}. The washout is in fact modulated by two parameters, $K_1$ and $z_*$. However, for large values of $m_1$, the parameter $K_1$ becomes huge and therefore, even if one has a large $z_*$, the frozen out asymmetry is completely washed out. On the other hand, when $z_*$ is small, e.g., $z_*=10^2$, the overall magnitude of $N_{B-L}$ decreases since $\beta\sim z_*^3 $. In this case however, $z_*^{-1}$ suppression in $\kappa$ is  not that significant compared to the previous one. Until $m_1\sim 10^{-2}$ eV, it shows the constant behaviour due to the mentioned nature of $f_\kappa$, however, at large $m_1$ values, it becomes strong enough to maintain the asymmetry in equilibrium for a period of time. The downfall is mostly dominated due  to this equilibrium asymmetry tracking and not due to the late time washout. 
\begin{figure}
\includegraphics[scale=.42]{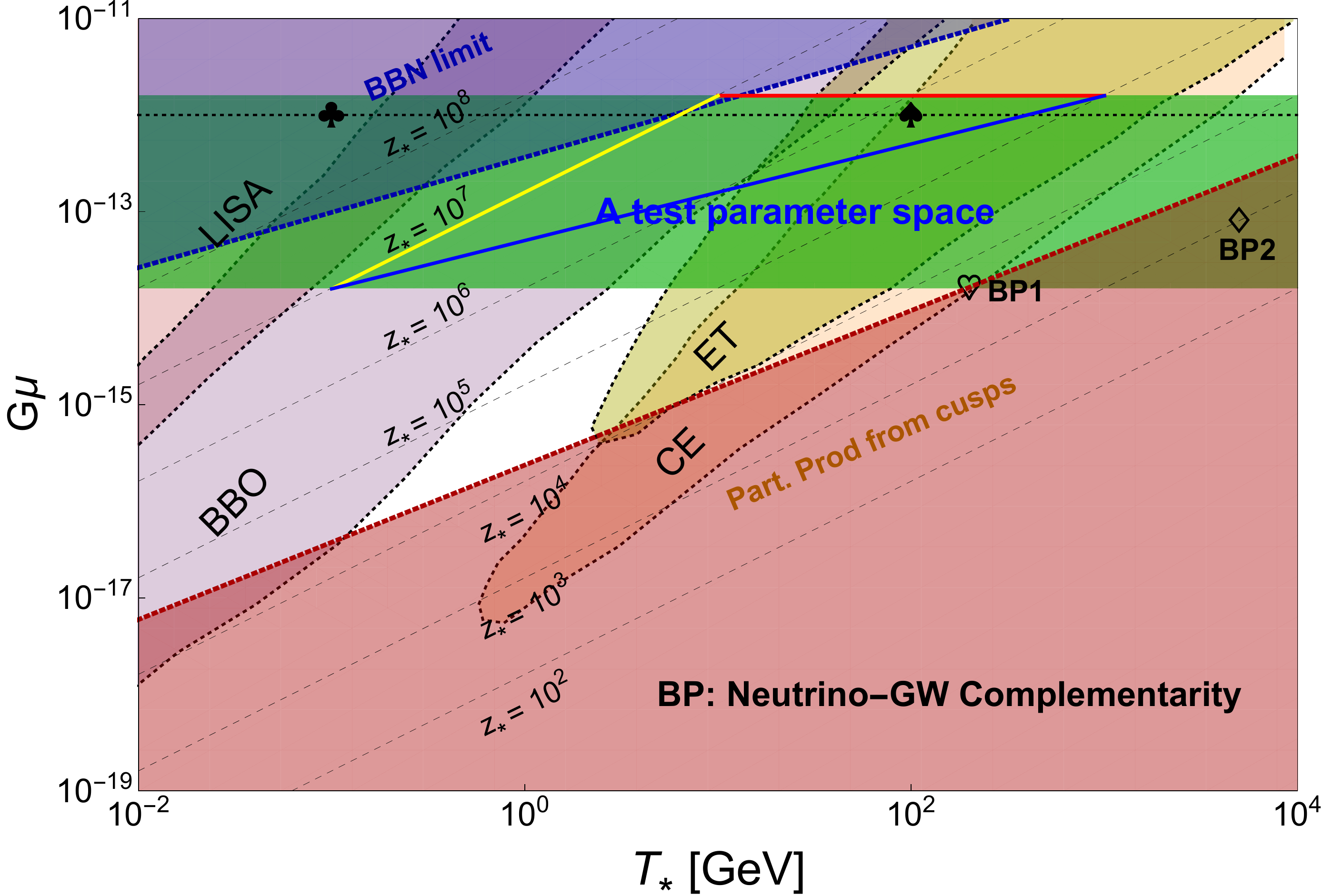} 
\caption{$G\mu$ vs. $T_*$ plot against the sensitivities of various GW detectors.}\label{fig2}
\end{figure}
Note that for $\omega<1/3$, for a fixed value of $z_*$, $\kappa$ increases (causes delayed freeze out and hence dilution of the asymmetry $N_{B-L}^{G0}$) and $\beta$ decreases (causes a decrease in $N_{B-L}^{eq}$). A concrete example is a matter domination, i.e.,  $\omega=0$, where $\kappa\sim \sqrt{z_*}$ and $\beta\sim z_*^{-3/2}$. Moreover, these kind of scenarios are inclusive of a late time entropy production which dilutes the produced asymmetry significantly\cite{matdom2,lepcs4}. Therefore, $\omega<1/3$ scenarios are utterly inefficient. This possibly strengthens the claim that in the future, should the  GW detectors find  a flat and then a rising signal, RH neutrino induced gravitational leptogenesis with a stiffer equation of state is a natural mechanism to associate with, since both of them, successful leptogenesis and the GW signal, are triggered by common theoretical ingredients.

\begin{figure}
\begin{center}
\includegraphics[scale=.65]{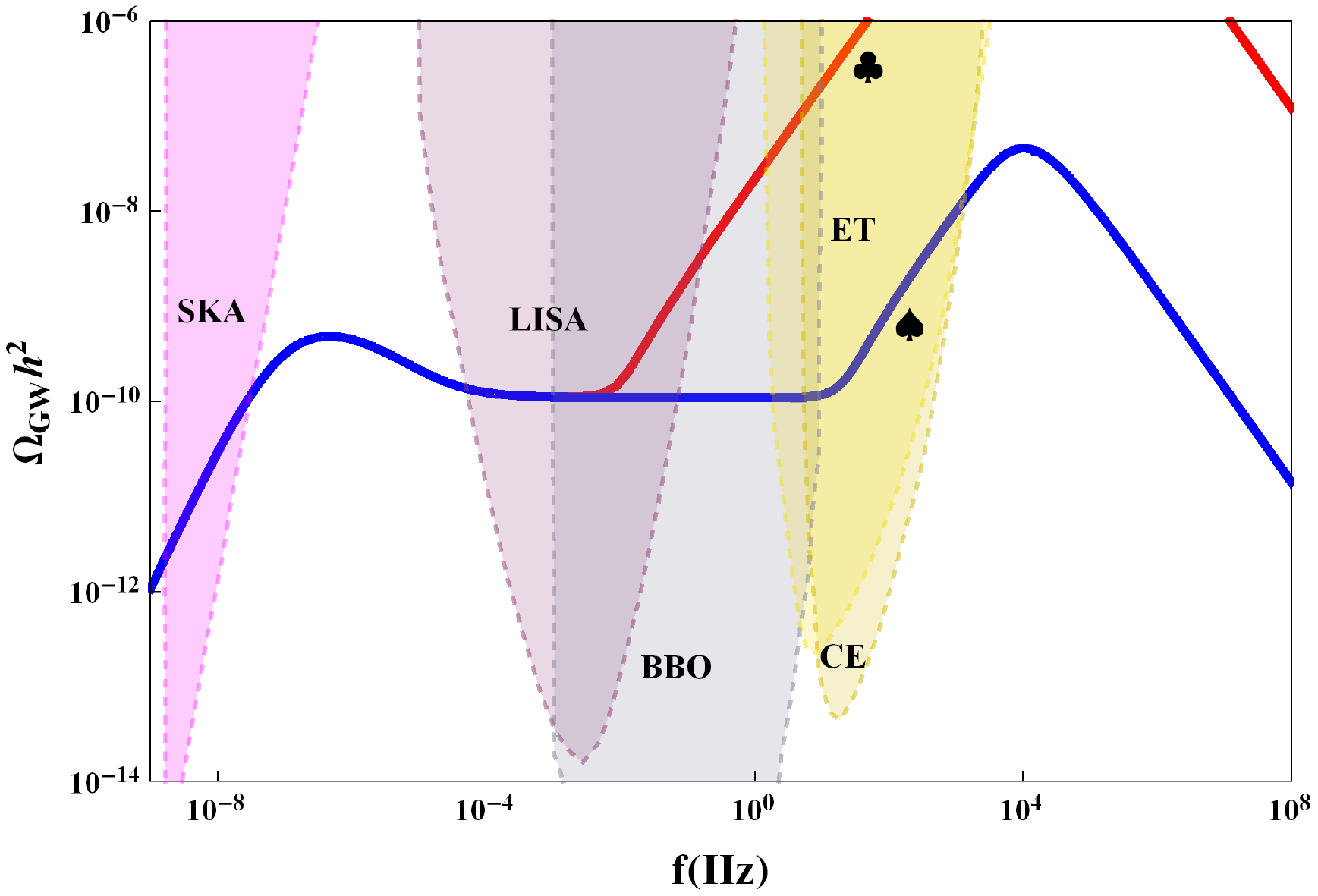} 
\caption{EOS: $\omega=1$. The curve in blue (red) is a valid (invalid) signal of leptogenesis. The curves are generated with $G\mu=10^{-12}$ and $T_*=10^{-1}$ GeV (red) and $T_*=10^{2}$ GeV (blue). A fall at a high frequency is due to the particle production from cusps for $l<l_{cric}=\frac{\mu^{-1/2}}{(\Gamma G\mu)^2}$\cite{spb2,partrad2}. We have shown the spectrum only for the fundamental mode.}\label{fig3}
\end{center}
\end{figure}

  In Fig.\ref{fig2}, we show the future sensitivities of the GW detectors such as LISA\cite{LISA}, BBO\cite{BBO}, CE\cite{CE}, ET\cite{ET} on the $G\mu-T_*$ plane. In the case of strong GW amplitudes, the most stringent constraint  comes from the effective number of neutrino species which reads $\int df f^{-1}\Omega_{GW}(f)h^2<5.6\times 10^{-6}\Delta N_{eff}$. Considering $\Delta N_{eff}\le 0.2$, the peak of the spectrum at $f_\Delta$, and taking into account contributions from the infinite number of modes that give a factor of $\zeta(7/3)$ amplification compared to the fundamental mode, the BBN constraint translates to $G\mu<T_*^{4/7}\left(1.72\times 10^{-22}\right)^{4/7}$. This has been shown by the blue exclusion region. On the other hand, to observe two spectral breaks (at $f_*$ and $f_\Delta$) distinctly, one should have $f_\Delta>f_*$ which translates to the constraint $G\mu>T_*^{4/5}\left(2.88\times 10^{-20}\right)^{4/5}$, where we consider particle production from cusps\cite{partrad2}. The corresponding  region has been shaded in red.  We have ignored the variation of the effective relativistic degrees of freedom even when $T_*$ is below the QCD phase transition. Proper temperature dependence of the same, would include a factor of 1.5-3 modification.  
Since we are entirely onto the gravitational leptogenesis (to motivate $\omega\neq 1/3$), we take $M^{max}_1\sim 10^8$ GeV so that the contribution from the decays are negligible. This gives an upper bound on the $T_{in}(\Lambda_{CS})$ that corresponds to $G\mu\lesssim 10^{-12}$. Therefore, the mechanism can be tested with reasonably strong GW amplitudes even for the flat part (Eq.\ref{flp1}).  For strong amplitudes, the spectral breaks are likely to happen at high-frequency GW detectors like CE and ET plus the bump like signals ($f_*$ and $f_\Delta$ are close to each other) in general lie outside those detectors.  In Fig.\ref{fig2}, the black point represented by $\spadesuit$ ($\clubsuit$), should (not) be a signal (see a supplementary Fig.\ref{fig3}). 

We shall end the discussion with a `Neutrino-Gravitational Waves Complementarity (NGWC)' or more generally, how this type of GW signal could be supplemented by low energy neutrino experiments.  NGWC depends on the $z_*$ and flavour structure of the theory or  more precisely, on the orthogonal matrix. From Fig.\ref{fig1}, it can be seen that, depending on the RH neutrino mass (hence $G\mu$), various $z_*$ values are sensitive to the neutrinoless double beta decay experiments (the $N_{B-L}$ curves intersect with the $N_{B-L}^{Obs}$ at the same time falls within the vertical green region). For the parameter set in Fig.\ref{fig1}, the NGWC points  fall  unfortunately  in the red region as well as they are well outside the GW detectors (showed by the  $\heartsuit$ and $\diamondsuit$ points in Fig.\ref{fig2}). However, if one decreases the $y_{ij}$, to produce correct $N_{B-L}$, for a fixed value of $G\mu$ one needs larger values of $z_*$, meaning the NGWC points would move towards the left side, i.e., towards the smaller values of $T_*$. The entire picture can be encapsulated within the triangle drawn on the test parameter space shaded in green in Fig.\ref{fig2}.  The red horizontal arm represents the constant $G\mu$ line along which the entries of $\Omega$ decrease as one goes from larger to smaller $T_*$. The yellow arm represents the constant $z_*$ line, as one goes along the line towards smaller $G\mu$ values, entries of $\Omega$ increase and the blue arm represents the constant (already predicted) orthogonal matrix and as one goes towards the higher values of $G\mu$, $z_*$ decreases or in other words, $T_*$ increases.   The blue arm is of great interest. If one has a completely determined orthogonal matrix, from Fig.\ref{fig1} the NGWC points can be determined with the sets of $M_1$ and $T_*$. This means the blue arm is a line of predictions  from the GW experiments, i.e., we can predict at which amplitude and at which frequency the spectral break would occur. The triangle as a whole can be pushed towards the larger $T_*$ values increasing $y_{ij}$. This implies, seesaw models which exhibit an orthogonal matrix with large imaginary part entries, would likely to show the spectral break at higher frequencies and therefore may not be tested with the planned detectors. These models are dubbed as `boosted' seesaw models where the light neutrino basis vectors and heavy neutrino basis vectors are strongly misaligned\cite{boost}. On the other hand, models with flavour structures close to `form dominance'\cite{Chen:2009um}  that typically predicts a real orthogonal matrix ($\Omega=P$, where $P$ is a permutation matrix), would show a spectral break within the frequency range of the current or planned GW detectors. \\

{\it Acknowledgements:} RS is supported  by the  MSCA-IF IV FZU - CZ.02.2.69/0.0/0.0/$20\_079$/0017754 project and acknowledges European Structural and Investment Fund and the Czech Ministry of Education, Youth and Sports. RS acknowledges Graham M. Shore and Pasquale Di Bari for an useful discussion on gravitational leptogenesis and boosted seesaw models respectively, Kai Schmitz for a helpful chat on Ref.\cite{lepcs3} and Sabir Ramazanov for discussions on cosmic strings in general.
{}
  \clearpage
  \section*{auxiliary}
\subsection*{AUX A: Flat plateau, loop number density normalisation and the turning point frequencies} 
 
{\textcolor{black}{\it A1. The standard expression:} }The normalised energy density parameter of gravitational waves at present time is expressed  as
\bea
\Omega_{GW}(t_0,f)=\frac{f}{\rho_c}\frac{d\rho_{GW}}{df}=\sum_k\Omega_{GW}^{(k)}(t_0,f).\label{feq}
\eea
The frequency derivative of  $\rho_{GW}$ is given by
\bea
\frac{d\rho_{GW}^{(k)}}{df}=\int_{t_F}^{t_0} \left[\frac{a(\tilde{t})}{a(t_0)}\right]^4 P_{GW}(\tilde{t},f_k)\frac{dF}{df}d\tilde{t},\label{a2}
\eea
where  $\frac{dF}{df}=f \left[\frac{a(t_0)}{a(\tilde{t})}\right]$ the quantity $P_{GW}(\tilde{t},f_k)$ represents the power emitted by the loops and is given by (see e.g., Ref.[68])
\bea
P_{GW}(\tilde{t},f_k)=G\mu^2\Gamma_k\int  n(l,\tilde{t}) \delta\left(f_k-\frac{2k}{l}\right) dl.\label{a3}
\eea
Integrating Eq.\ref{a3} over the loop lengths gives 
\bea
P_{GW}(\tilde{t},f_k)=\frac{2kG\mu^2 \Gamma_k}{f_k^2} n(\tilde{t},f_k)=\frac{2kG\mu^2 \Gamma_k}{f^2\left[\frac{a(t_0)}{a(\tilde{t})}\right]^2}n\left(\tilde{t},\frac{2k}{f}\left[\frac{a(\tilde{t})}{a(t_0)}\right]\right).\label{a4}
\eea
From Eq.\ref{a4} and Eq.\ref{a2} one gets
\bea
\frac{d\rho_{GW}^{(k)}}{df}=\frac{2kG\mu^2 \Gamma_k}{f^2}\int_{t_{osc}}^{t_0} \left[\frac{a(\tilde{t})}{a(t_0)}\right]^5 n\left(\tilde{t},\frac{2k}{f}\left[\frac{a(\tilde{t})}{a(t_0)}\right]\right)d\tilde{t}
\eea
and therefore the energy density corresponding to the mode `$k$' is given by
\bea
\Omega_{GW}^{(k)}(t_0,f)=\frac{2kG\mu^2 \Gamma_k}{f\rho_c}\int_{t_{osc}}^{t_0} \left[\frac{a(\tilde{t})}{a(t_0)}\right]^5 n\left(\tilde{t},\frac{2k}{f}\left[\frac{a(\tilde{t})}{a(t_0)}\right]\right)d\tilde{t}.\label{form1}
\eea

Using the VOS equations and considering the loop production function as a delta function (see, e.g., Ref.[74]), it is easy to obtain the most general formula for the number density in an expanding background that scales as $a\sim t^\beta$. The expression is given by
\bea
n(\tilde{t},l_{k}(\tilde{t}))=\frac{A_\beta}{\alpha}\frac{(\alpha+\Gamma G \mu)^{3(1-\beta)}}{\left[l_k(\tilde{t})+\Gamma G \mu\tilde{t}\right]^{4-3\beta}\tilde{t}^{3\beta}}.\label{genn}
\eea
The Eq.\ref{form1} can be expressed in the conventional form that are used in many papers (e.g., Ref.[39,40]) using the time dependence of the loop length which gives  initial time $t_i^{(k)}$ as
\bea
t_i^{(k)}=\frac{l_k(\tilde{t})+\Gamma G\mu \tilde{t}}{\alpha+\Gamma G\mu},\label{tin}
\eea
Now using Eq.\ref{tin}, the number density in Eq.\ref{genn} can be re-expressed as
\bea
n(\tilde{t},l_{k}(\tilde{t}))=\frac{A_\beta(t_i^{(k)})}{\alpha(\alpha+\Gamma G \mu)t_i^{(k)4}}\left[\frac{a(t_i^{k})}{a(\tilde{t})}\right]^3.\label{init}
\eea
Putting the value of $n(\tilde{t},l_{k}(\tilde{t}))$ from Eq.\ref{init} into Eq.\ref{form1}, one gets the standard expression
\bea
\Omega_{GW}^{(k)}(t_0,f)=\frac{2kG\mu^2 \Gamma_k}{f\rho_c \alpha(\alpha+\Gamma G \mu)}\int_{t_{osc}}^{t_0} \left[\frac{a(\tilde{t})}{a(t_0)}\right]^5\frac{C_{\rm eff}(t_i^{(k)})}{t_i^{(k)4}} \left[\frac{a(t_i^{k})}{a(\tilde{t})}\right]^3 d\tilde{t},\label{form2}
\eea
where we have renamed $A_\beta$ as $C_{\rm eff}$.\\

{\textcolor{black}{\it A2. The flat plateau:} }To obtain the GW spectrum from the loops that are produced and decay during the radiation domination,  it  is convenient to do the integration in Eq.\ref{form1}  with respect to the scale factor which reads
\bea
\Omega_{GW}^{(k)}(t_0,f)=\frac{16\pi}{3\zeta(\delta)}\left(\frac{G\mu}{H_0}\right)^2\frac{\Gamma}{f a(t_0)}\int_{a_*}^{a_{eq}}H(a)^{-1} \left[\frac{a(\tilde{t})}{a(t_0)}\right]^4 n\left(\tilde{t},\frac{2k}{f}\left[\frac{a(\tilde{t})}{a(t_0)}\right]\right)da,\label{form3}
\eea
where
\bea
H=H_0\Omega_r^{1/2}\left(\frac{a(\tilde{t})}{a(t_0)}\right)^{-2}~~{\rm with}~~\Omega_r\simeq 9\times10^{-5}.
\eea
The number density $n\left(\tilde{t},l_k(\tilde{t})\equiv \frac{2k}{f}\left[\frac{a(\tilde{t})}{a(t_0)}\right]\right)$ in Eq.\ref{genn} (in radiation domination) can also be expressed in terms of the scale factor as
\bea
n(\tilde{t},l_{k}(\tilde{t}))=\frac{A_r}{\alpha}\frac{(\alpha+\Gamma G \mu)^{3/2}}{\left[\frac{2}{f}\left[\frac{a(\tilde{t})}{a(t_0)}\right]+\Gamma G \mu/2H\right]^{5/2}(2H)^{-3/2}.}\label{numrad}
\eea
Putting Eq.\ref{numrad} in Eq.\ref{form3} and after performing the integration one gets 
\bea
\Omega_{GW}^{(1)}(f)=\frac{128\pi G\mu}{9\zeta(\delta)}\frac{A_r}{\epsilon_r}\Omega_r(1+\epsilon_r)^{3/2}\left[\left( \frac{f}{f+\epsilon_r f_{\rm min}\left(\frac{t_*}{t_{\rm eq}}\right)^{1/2}}\right)^{3/2}-\left( \frac{f}{f+\epsilon_r f_{\rm min}}\right)^{3/2}\right],
\eea
where we define $\epsilon_r=\alpha/\Gamma G \mu$ the $f_{min}=\frac{2}{\alpha t_*}\frac{a_*}{a_0}=\frac{4H_0\Omega_r^{1/2}}{\alpha}\frac{a_0}{a_*}$ is the minimum frequency emitted by  a given loop.  Given the scaling solution of the loop production rate, which decreases with the fourth power in time,  $f\simeq f_{\rm min}$ is  a reasonable assumption. Then, with $t_i\ll t_{\rm eq}$ one has 
\bea
\Omega_{GW}^{(1)}(f)=\frac{128\pi G\mu}{9\zeta(\delta)}\frac{A_r}{\epsilon_r}\Omega_r\left[(1+\epsilon_r)^{3/2}-1\right]. \label{flp}
\eea
The expression for the flat plateau matches with Ref.[74] barring the factor $\zeta(\delta)$ in the denominator. This is due to the fact that  definition of the $\Omega_{GW}^{(k)}(f)$ in Eq.\ref{form1} is inclusive of $\Gamma^{k}$. \\

{\textcolor{black}{\it A3. The turning point frequencies:}} In the above, it is assumed that the dominant emission comes from the very earliest epoch of loop creation.  Nonetheless, a precise value of the time can be calculated by maximizing the integral in Eq.\ref{form3} with respect to $\tilde{t}$ which gives 
\bea
\tilde{t}_M\simeq\frac{2}{f\Gamma G\mu}\frac{a_M}{a_0}\equiv \frac{1}{2\Gamma G\mu}\frac{4 a_M}{f a_0}\equiv \frac{l_i}{2\Gamma G\mu},\label{half}
\eea
where $f$ is the frequency observed today which was emitted at time $\tilde{t}_M$ when the a given initial loop $l_i=\alpha t_i$ reached to the half of its size $l_i/2$, i.e., $\tilde{t}_M$ is eventually the half-life of the loop. If time $t_*$ at which the most recent radiation domination begins, an approximate  frequency up to which the spectrum shows a flat plateau is given by
\bea
f_*=\sqrt{\frac{8}{\alpha\Gamma G\mu}}t_*^{-1/2}t_0^{-2/3}t_{\rm eq}^{1/6}\simeq \sqrt{\frac{8 z_{\rm eq}}{\alpha\Gamma G\mu}}\left(\frac{t_{\rm eq}}{t_*}\right)^{1/2}t_0^{-1}.\label{br0}
\eea
Similarly, using the critical length $l_{cric}=\frac{\mu^{-1/2}}{(\Gamma G \mu)^2}$ for cusp like structures,  the second turning point frequency can be computed as
\bea
f_\Delta\simeq 9\sqrt{\alpha}\left(G\mu\right)^{5/4} \left(\frac{M_{pl}}{T_*}\right)f_*,\label{turn1st}
\eea
where we have assumed that  $f_\Delta/f_* \simeq\sqrt{t_*/t_\Delta}$ (cf. Eq.\ref{turn1st}), and for simplicity we consider $g_*(T)=g_*\simeq 106$ throughout. Therefore, for post QCD phase transition $T\lesssim 200$ MeV, the formula is bit errorful.\\
To observe both the frequencies distinctively, one should have $f_\Delta>f_*$. This gives the following restriction on the parameter space
\bea
G\mu>T_*^{4/5}\left(2.88\times 10^{-20}\right)^{4/5}\label{const1}
\eea
which is shown by the red region in Fig.2. \\

{\textcolor{black}{\it A4. The BBN limit:}} To be consistent with the the number of effective neutrino species, the GW energy density has to comply with 
\bea
\int_{f_{BBN}}^{f_{max}}\frac{df}{f}\Omega_{GW}h^2<5.6\times 10^{-6}\Delta N_{eff},
\eea
with $\Delta N_{eff}<0.2$. Considering  the dominant contribution from the non-flat part after the first turning point frequency $f_*$, the following constraint on the parameter space can be obtained 
\bea
G\mu<T_*^{4/7}\left(1.22\times 10^{-22}\right)^{4/7}\label{const2}
\eea
which is shown in  the blue region in Fig.2. The constraints in Eq.\ref{const1} and Eq.\ref{const2} are derived for $\alpha=0.1$.\\

{\textcolor{black}{\it A5. Numerical simulation  vs. VOS model loop number density and the normalisation:}} The number density obtained from numerical simulation is given by (see, Ref.[68]) (considering the loops created during radiation domination)
\bea
n(\tilde{t},l_{k}(\tilde{t}))=\frac{0.18}{\left[l_k(\tilde{t})+\Gamma G \mu\tilde{t}\right]^{5/2}\tilde{t}^{3/2}}.\label{nub}
\eea
On the other hand considering the analytic approach, i.e., using Velocity dependent One Scale (VOS) the same is obtained as
\bea
n(\tilde{t},l_{k}(\tilde{t}))=\frac{A_r}{\alpha}\frac{(\alpha+\Gamma G \mu)^{3/2}}{\left[l_k(\tilde{t})+\Gamma G \mu\tilde{t}\right]^{5/2}\tilde{t}^{3/2}}\equiv\frac{A_r N_\alpha}{\left[l_k(\tilde{t})+\Gamma G \mu\tilde{t}\right]^{5/2}\tilde{t}^{3/2}},\label{vos}
\eea
where $A_r=5.4$. As mentioned before, the  VOS model assumes all the loops are of same length at creation. However, at the moment of creation, the loops may follow a distribution depending on $\alpha$. If so, the above formula should be modified as 
\bea
n(\tilde{t},l_{k}(\tilde{t}))=\frac{A_r \int w(\alpha) N_\alpha d\alpha}{\left[l_k(\tilde{t})+\Gamma G \mu\tilde{t}\right]^{5/2}\tilde{t}^{3/2}}.
\eea
Therefore, to the make VOS formula in Eq.\ref{vos}  consistent with the numerical result, one has to normalise Eq.\ref{vos}, i.e., 
\bea
n(\tilde{t},l_{k}(\tilde{t}))=\frac{\mathcal{F}_\alpha A_r N_\alpha}{\left[l_k(\tilde{t})+\Gamma G \mu\tilde{t}\right]^{5/2}\tilde{t}^{3/2}},~~{\rm with}~~\mathcal{F}_\alpha =N_\alpha ^{-1}\int w(\alpha) N_\alpha d\alpha
\label{mvos}
\eea
As one can see that for $\alpha=0.1$, Eq.\ref{nub} and Eq.\ref{vos} is consistent for $\mathcal{F}_\alpha\sim 0.18/( A_r \sqrt{\alpha})\sim 0.1$.\\

{\textcolor{black}{\it A6. The spectral shape beyond the first turning point frequency:}} As mentioned previously in the main text, when the number of modes increases in the sum, the spectral behaviour  beyond the turning point deviates from that of the fundamental mode (see e.g., Ref.[29,40] ). The reason being the following: \\

From Eq.\ref{form2} it is evident that 
\bea
\Omega_{GW}(f)=\sum_k\Omega_{GW}^{(k)}(f)=\sum_k k^{-\delta}\Omega^{(1)}(f/k).\label{sumk}
\eea
Now to perform the sum one can expand the RHS of Eq.\ref{sumk} for some first few benchmark modes, i.e.,
\bea
\Omega_{GW}(f)&=&\sum_k k^{-\delta}\Omega^{(1)}(f/k) \nonumber \\ &=& 1^{-\delta}\Omega^{(1)}(f/1)+ m^{-\delta}\Omega^{(1)}(f/m)+ n^{-\delta}\Omega^{(1)}(f/n)+ r^{-\delta}\Omega^{(1)}(f/r)+...,
\eea
where the integers obey $1<m<n<r$. This suggests, if one keeps on increasing the mode numbers, there should be a critical value  $k\equiv k_*$ for which the amplitude $\Omega_{GW}^{(1)}(f_*=f/k_*)$ contributes to the frequency $f$. Therefore, the  sum can be split  into two parts. The first one is from $k=1$ up to $k_*$ for which the the amplitude at $f$ receives contributions from the non-flat part  and the second one is from $k_*$ to $k_{max}$ for which the test point receives contribution from the flat part, i.e., 
\bea
\Omega_{GW}(f)&=&\sum_{k=1}^{k=k_{*}}k^{-\delta}\Omega_{GW}^{(1)}(f/k>f_*)+\sum_{k=k_{*}}^{k=k_{max}}k^{-\delta}\Omega_{GW}^{(1)}(f/k<f_*)\\
&=&\sum_{k=1}^{k=k_{*}}k^{-\delta}\Omega_{GW}^{\rm plt}\left(\frac{f/k}{f_*}\right)+\sum_{k=k_{*}}^{k=k_{max}}k^{-\delta}\Omega_{GW}^{\rm plt}.\label{2sum}
\eea
The first term  in Eq.\ref{2sum} gives the dominant contribution. In the large $k_*$ limit, the sum is therefore
\bea
\Omega_{GW}(f)\simeq  \Omega_{GW}^{\rm plt} \left(\frac{f}{f_*} \right)\sum_{k=1}^{k=k_{*}}k^{-(\delta +1)}\simeq \Omega_{GW}^{\rm plt} \left(\frac{f}{f_*} \right)\zeta(\delta + 1).
\eea
Therefore,  for an equation of states like kination, the spectral shape is quite similar to the $k=1$ mode even after adding the contributions from the larger number of modes.\\ 

%{\textcolor{black}{\it A7. A $\Omega_{GW}$ vs. $f$ plot for two bench mark scenarios represented by $\spadesuit$ and  $\clubsuit$ in the main text.}}

\subsection*{AUX B: Evolution of the lepton asymmetry and derivation of the master equation} 

The energy density in  a general equation of state red-shifts as
$\rho_\omega \propto  a^{-3(1+\omega)}$. We assume there is no further entropy production after the instantaneous reheating. Therefore the scale factor is inversely proportional to the temperature, i.e., $\rho_\omega \propto  T^{3(1+\omega)}$. Since the energy density of radiation and field $\phi_\omega$ should be equal at the critical temperature $T_*$, the proportionality constant $\sigma_\omega$ can then be obtained as
\bea
\sigma_\omega=\sigma_{\rm rad} T_*^{1-3 \omega},
\eea
where the energy density in radiation domination is given by $\rho_{\rm rad}=\sigma_{\rm rad}T^4\equiv (\pi^2 g_*/30)T^4$. The total energy  at an arbitrary temperature $T$ is then given by
\bea
\rho(z)=\rho_{\rm rad}(z)\left[1+\left(\frac{z}{z_*}\right)^{1-3\omega}\right],
\eea
where we define $z_*(z)=M_1/T_*(M_1/T)$. The  modified Hubble parameter and the $\dot{R}$ in the general equation of state are then given by
\bea
H_\omega(z)=H_{\rm rad}(z)\left[1+\left(\frac{z}{z_*}\right)^{1-3\omega}\right]^{1/2}, \label{hubw}\\
\dot{R}=\sqrt{3}\sigma_{\rm rad}^{3/2}(1-3\omega)(1+w)\frac{T^6}{\tilde{M}_{Pl}^3}\left(\frac{z}{z_*}\right)^{\frac{3}{
2}(1-3\omega)}.\label{rw}
\eea
Given the Hubble parameter in Eq.\ref{hubw}, the expression for the lepton number violating interactions $W(z)\equiv\Gamma_{\Delta L=2}/Hz$ can be generalised as
\bea
W_\omega=W_{\rm rad}\left[1+\left(\frac{z}{z_*}\right)^{1-3\omega}\right]^{-1/2}.\label{washout}
\eea
The most general Boltzmann equations (BEs) for leptogenesis with seesaw Lagrangian minimally coupled to gravity are
\bea
\frac{dN_{N_1}}{dz}&=&-D\left[N_{N_1}-N_{N_1}^{eq}\right],\label{rhp}\\
\frac{{\bf d  N_{B-L}}}{\bf dz}&=&-D\varepsilon_1\left[N_{N_1}-N_{N_1}^{eq}\right]-({\bf W_{\Delta L=2}+W_{\rm \bf ID}})\left[{\bf N_{B-L}}-{\bf N_{B-L}^{eq}}\right],\label{asp}
\eea
where the first equation governs the production of RH neutrinos and the first term in the second equation represents the contribution to the lepton asymmetry  from RH neutrino  decays. Since we are neglecting the contribution from decays, only the  second equation with the terms in `bold' is relevant. Note that recently in Ref.[44], another curvature-induced evolution term that modulates of the asymmetry production dynamics at ultra-high temperatures has been introduced. We neglect that term in our computation. However, that will not change the qualitative features of our final results.
To obtain a simpler form of the Boltzmann equation it is convenient to simplify the expression of the equilibrium asymmetry and the lepton number violating processes. Using the orthogonal parametrisation of the Dirac neutrino mass matrix $m_D=U\sqrt{m}\Omega\sqrt{M}$, the equilibrium asymmetry can be expressed as a power law in $z$ as
\bea
N_{B-L}^{eq}=\frac{\beta}{z^q} ~~{\rm with} ~~q=\frac{7+9 \omega}{2}.\label{qv}
\eea
Here the parameter $\beta$ is given by 
\bea
\beta=\frac{\sqrt{3}\pi^2 }{36 (4\pi v)^4}\sigma_{\rm rad}^{3/2}(1-3\omega)(1+\omega)\frac{M_1^5}{\tilde{M}_{pl}^3}\mathcal{Y},
\eea
 where the parameter $\mathcal{Y}$ encodes CP violation in the theory and is given by
 \bea
  \mathcal{Y}=
\sum_{j>i}\frac{\sum_{k,k^\prime} m_km_{k^\prime} {\rm Im\left[ \Omega_{ki}^*\Omega_{kj}\Omega_{k^\prime i}^*\Omega_{k^\prime j}\right]}}{\xi(3) } {\rm ln \left(\frac{M_j^2}{M_i^2}\right)}z_*^{\frac{3}{2}(3\omega-1)}.\label{ypara}
\eea
\begin{figure}
\includegraphics[scale=.49]{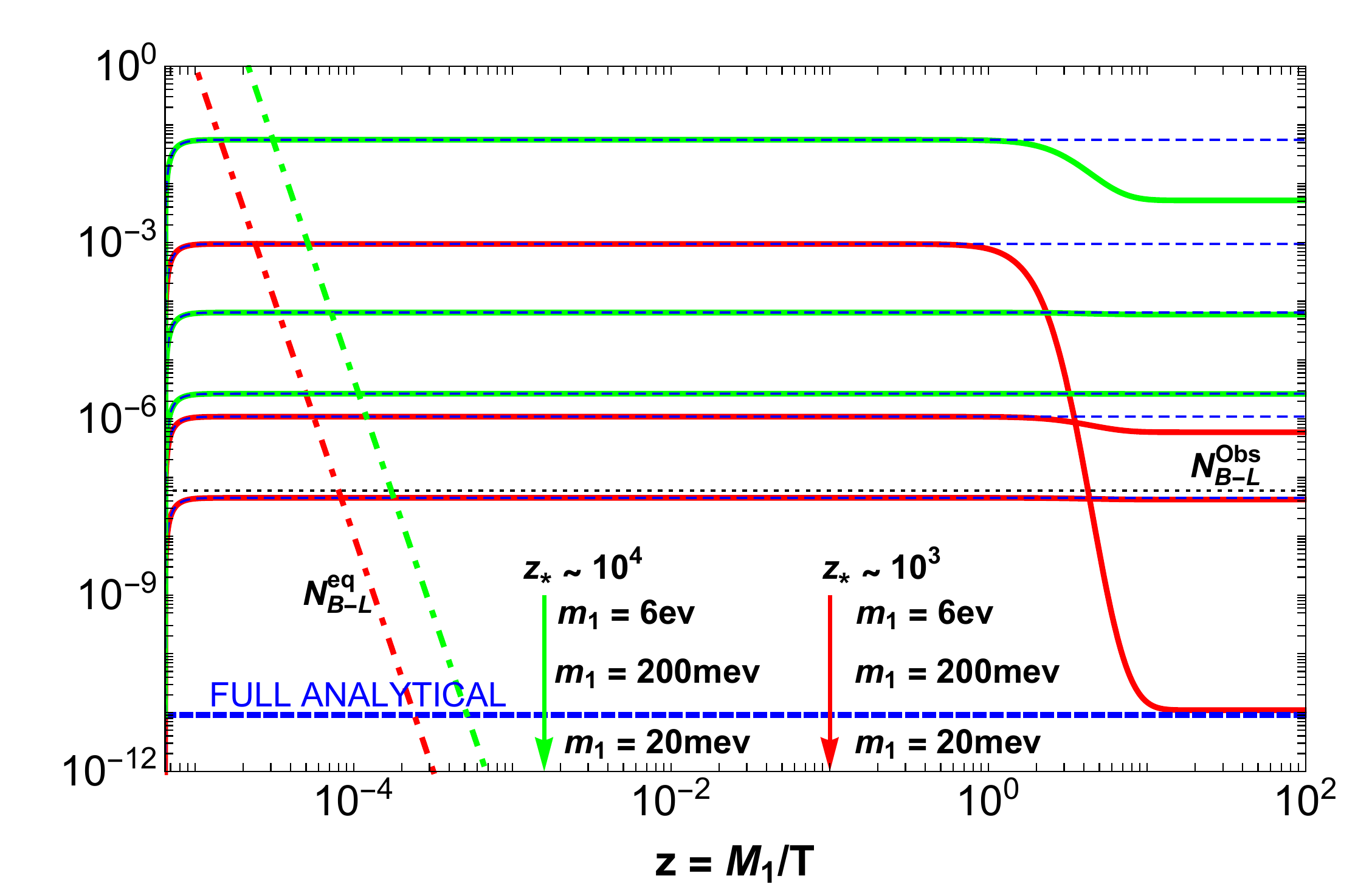}\\ \includegraphics[scale=.49]{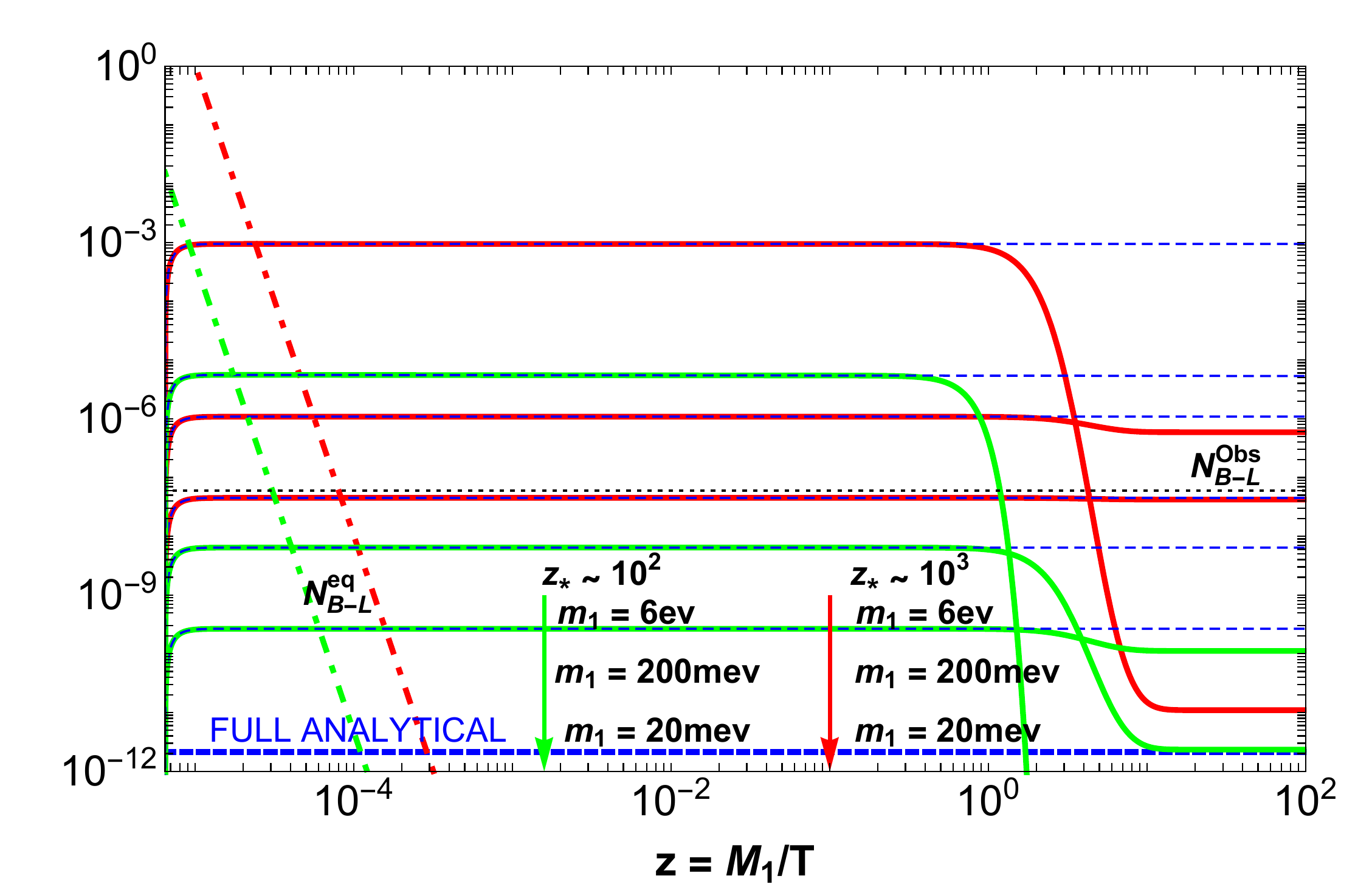}
\caption{Top: EOS: $\omega=1$. Evolution of the gravitationally produced asymmetry for different values of $m_1$. We have taken $M_1=10^8$ GeV, $z_*=1.3 \times 10^3$ (red), $10^4$ (green) as benchmark values.  Bottom: EOS: $\omega=1$. Evolution of the gravitationally produced asymmetry for different values of $m_1$. We have taken $M_1=10^8$ GeV, $z_*= 1.3\times 10^3$ (red), $10^2$ (green)  as  benchmark values. For both the plots we use $M_3=10^{14}$ GeV, $M_2=10^{12}$ GeV, $x_{ij}=\pi/4$, $y_{ij}=10^{-4}$, $\Delta m_{12}^2=7.4\times 10^{-5} {\rm eV^2}$, $\Delta m_{32}^2=2.4\times 10^{-3} {\rm eV^2}$ and  $z_{\rm in}=\sqrt{M_1/\tilde{M}_{pl}}$. The thick blue dashed line shows a match bewteen numerical solutions and the master equation obtained in Eq.\ref{fulana}. The thin blue dashed lines show a match bewteen numerical solutions and  solution obtained (without the late time $N_1$-washout) in Eq.\ref{noinv}.}\label{fig4}
\end{figure}
In the $z_*\gg z$ limit, two relevant lepton number violating processes, i.e., $W_{\Delta L=2}$ scattering and the inverse decay $W_{\rm ID}$ can be obtained as 
\bea
W_{\Delta L=2}(z)=\frac{\kappa}{z^p}~~{\rm with} ~~p=\frac{5-3\omega}{2},\label{pv}
\eea
where $\kappa$ depends on the $M_1$, the light neutrino masses and $z_*$ as 
\bea
\kappa=
\frac{12m^* M_1}{\pi^2 v^2 }\left( \left[\frac{\sqrt{\sum_i m_i^2}}{m^*}\right]^2+K_1^2-\frac{2m_1^2}{m^{*2}}\right)z_*^{\frac{1}{2}(1-3\omega)}\label{kapa}
\eea
and 
\bea
W_{\rm ID}(z)\simeq \frac{1}{4}K_1z_*^{-\frac{1}{2}(3\omega-1)}z^{\frac{7+3\omega}{2}-1}\mathcal{K}_1(z).\label{invde}
\eea
As mentioned earlier, at very high temperature the inverse decays are negligible. Therefore one can simply solve the BE
\bea
\frac{{ d  N_{B-L}}}{ dz}=-\frac{\kappa}{z^p}\left[N_{B-L}-\frac{\beta}{z^q}\right]\label{begw2}
\eea
to obtain the an expression for the frozen out asymmetry $N_{B-L}^{G0}$. Then the final asymmetry can be obtained as
\bea
N_{B-L}^f=N_{B-L}^{G0}e^{-\int_0^\infty W_{\rm ID}  (z) dz}.
\eea
For $\omega=1$, the solution for $N_{B-L}^{G0}$ is obtained as
\bea
N_{B-L}^{G0}=\frac{\kappa\beta}{\kappa-8}\left(z^{-8}-z_{in}^{\kappa-8}z^{-\kappa}\right)
\eea
which in the small $\kappa$ and $z\gg z_{\rm in}$ limit simplifies as 
\bea
N_{B-L}^{G0}=\frac{\kappa\beta}{8z_{\rm in}^8}.\label{noinv}
\eea

The washout by the inverse decays can be obtained using Eq.\ref{invde} and integral properties of the Bessel function $\mathcal{K}_n(z)$ 
\bea
\int_0^\infty z^{\alpha-1}\mathcal{K}_n(z)dz=2^{\alpha-2}\Gamma\left[\frac{\alpha-n}{2}\right]\Gamma\left[\frac{\alpha+n}{2}\right].
\eea
 The washout factor comes out as
\bea
e^{-\int_0^\infty W_{\rm ID}  (z) dz}\equiv \mathcal{W}_{W_{\rm ID}(K_1,z_*,\omega=1)}={\rm Exp}\left[-\frac{4K_1}{z_*}\right].
\eea
Therefore the master formula for the final asymmetry that can be used for a numerical scan is given by
\bea
N_{B-L}^f\simeq\frac{\kappa\beta}{8z_{\rm in}^8}{\rm Exp}\left[-\frac{4K_1}{z_*}\right].\label{fulana}
\eea
which very accurately reproduces the numerical result as shown in Fig.4 with the phrase ``FULL ANALYTICAL".

\end{document}